\documentclass[letterpaper,12pt,twocolumn]{article}

\usepackage[utf8]{inputenc}
\usepackage{graphicx}
\usepackage{amsmath}
\usepackage{amssymb}
\usepackage{geometry}
\usepackage{hyperref}
\usepackage{svg}
\usepackage{comment}
\usepackage{float}
\usepackage[font=small,skip=0pt]{caption}
\usepackage{biblatex}
\usepackage[para,bottom]{footmisc}

\newcommand{\ls}{\begin{itemize}\item}

\renewcommand{\le}{\end{itemize}}

\newcommand{\es}{\begin{enumerate}\item}

\newcommand{\ee}{\end{enumerate}}
\newcommand{\Latex}{\LaTeX\ }

\title{Litrepl: Literate Paper Processor Promoting Transparency More Than Reproducibility}
\author{Sergei Mironov \\
        \texttt{sergei.v.mironov@proton.me}}
\date{\today}

\begin{document}

\maketitle

\begin{abstract}

Litrepl is a lightweight text processing tool designed to recognize and evaluate
code sections within Markdown or \Latex documents. This functionality is useful
for both batch document section evaluation and interactive coding within a text
editor, provided a straightforward integration is established. Inspired by
Project Jupyter, Litrepl aims to facilitate the creation of research documents.
In the light of recent developments in software deployment, however, we have
shifted our focus from informal reproducibility to enhancing transparency in
communication with programming language interpreters, by either eliminating or
clearly exposing mutable states within the communication process.

\end{abstract}

\section{Statement of Need}

\begin{figure*}[!hbt]
  \centering
  \includegraphics{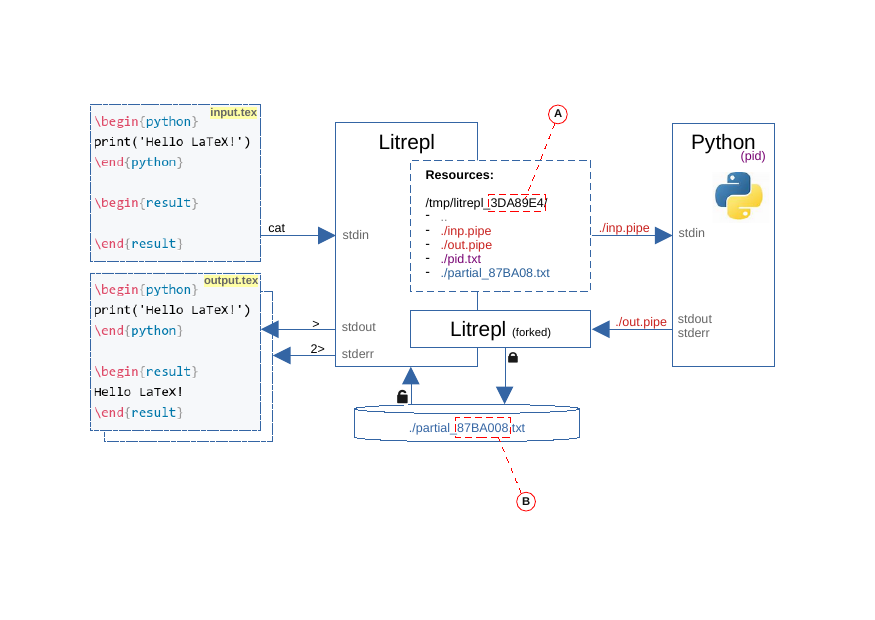}
  \caption{Litrepl resource allocation diagram. Hash \textbf{A} is computed
  based on the Litrepl working directory and the interpreter class. Hash
  \textbf{B} is computed based on the contents of the code section.}
  \label{fig:resource-allocation}
\end{figure*}

The concept of \textit{Literate Programming} was formulated by Donald Knuth,
suggesting a shift in focus from writing code to explaining to human beings what
we want a computer to do. This approach is embodied in the WEB
system\cite{Knuth1984lp} and its descendant family of tools, whose name refers
to a text document format containing the "network" of code sections interleaved
with text.

The system could both render such text into human-readable documentation and
compile machine-executable code. Over time, this concept has evolved, showing a
trend towards simplification (\textcite{Ramsey1994lps}).

Concurrently, a concept of human-computer interaction often called the
\textit{Read-Evaluate-Print Loop} or ''REPL'' gained traction, notably within
the LISP an APL communities (\textcite{Spence1975apl},
\textcite{McCarthy1959recfun}, \textcite{Iverson1962apl}).

The combination of a command-line interface and a language interpreter enables
incremental and interactive programming, allowing users to directly modify the
interpreter state. By maintaining human involvement in the loop, this approach
is believed to facilitate human thought processes (\textcite{Granger2021litcomp}).

A significant milestone in this field was the IPython interpreter
(\textcite{Perez2007IPython}), which later evolved into the Jupyter Project. Its
creators introduced a new document format called the Jupyter Notebook
(\textcite{Kluyver2016jupnb}), characterized by a series of logical sections of
various types, including text and code, which could directly interact with
programming language interpreters. This interactive communication, akin to REPL
style programming, allows the creation of well-structured documents suitable for
presentations and sharing. The concept underpinning these developments is termed
\textit{Literate Computing}\cite{Perez2015blog}, which includes goals of
spanning a wide audience range, boosting reproducibility, and fostering
collaboration. To achieve these objectives, several technical decisions were
made, notably the introduction of bidirectional communication between the
computational core, known as the Jupyter Kernel, and the Notebook web-based
renderer, along with another layer of client-server communication between the
web-server and the user’s web browser.

While we recognize the importance of all goals within the Literate Computing
framework, we think that the goal of reproducibility is more important than
others. Addressing it alone would suffice to enhance communication among
time-separated and space-separated researchers and significantly expand the
audience. However, as it became clear (\textcite{Dolstra2010}), this challenge
extends beyond the scope of a single human-computer interaction system, and even
beyond the typical boundaries of software distribution management for a
particular programming language. A comprehensive solution to the software
deployment problem operates at the entire operating system level.

Following \textcite{Vallet2022}, we suggest changing the focus of human-computer
interaction towards simplicity and transparency. We saw an opportunity to
implement a tool that would offer REPL-style editing, be compatible with
existing code editors and text formats, thus maintaining familiar editing
practices, contain only a few hidden state variables, and have a significantly
smaller codebase.

We introduce
\textit{Litrepl}\footnote{\url{https://github.com/sergei-mironov/litrepl}},
a text processor that employs the following approaches: first, utilizing
straightforward bidirectional text streams for inter-process communication with
language interpreters to evaluate code; second, advocating for the reuse of
existing text document formats. In both the Markdown and \Latex evaluators we
have implemented, simplified parsers are used to distinguish code and result
sections from the rest of the document.  As of now, we support Python and Shell
interpreter families, as well as a custom large language model communication
interpreter. Finally, we strive to leverage POSIX\cite{POSIX2024} operating
system facilities as much as possible.

\section{How it works}

Litrepl is implemented as a command-line text utility. Its primary function is
to take a text document as input, process it according to specified command-line
arguments and environment settings, and then output the resulting document
through its standard output.

The operation of Litrepl is best illustrated through the example below. Consider
the document named \verb|input.tex|:

\begin{verbatim}
$ cat input.tex
\begin{python}
import sys
print(f"I use {sys.platform} btw!")
\end{python}
\begin{result}
\end{result}
\end{verbatim}

This document contains a Python code section and an empty result section marked
with the corresponding \Latex environment tags. To "execute" the document we
pipe it though the Litrepl processor as follows:

\begin{verbatim}
$ cat input.tex | litrepl
\begin{python}
import sys
print(f"I use {sys.platform} btw!")
\end{python}
\begin{result}
I use linux btw!
\end{result}
\end{verbatim}

Now we can see the expected statement about the author's operating system.  The
side-effect of this execution is the started session of the python interpreter
which is now running in the background. We can modify its state by adding more
section to the document and executing them selectively or e.g. by accessing
\verb|litrepl repl python| terminal. So for example, setting \verb|sys.platform|
to another value and re-evaluating the document would yield a different
statement.

\subsection{Interfacing Interpreters}

Litrepl communicates with interpreters using two uni-directional text streams:
one for writing input and another for reading outputs. To establish effective
communication, the interpreter should conform to the following general
assumptions:

\begin{itemize}
  \item Synchronous single-user mode, which is implemented in most interpreters.
  \item A capability to disable command line prompts. Litrepl relies on the echo
        response, as described below, rather than on prompt detection.
  \item The presence of an echo command or equivalent. The interpreter must be
        able to echo an argument string provided by the user in response to the
        echo command.
\end{itemize}

In Litrepl, these details are hardcoded for several prominent interpreter
families, which we refer to as \textit{interpreter classes}. At the time of
writing, Litrepl supports three such classes: \texttt{python}, \texttt{sh}, and
\texttt{ai}. Using these names in command line arguments, users can configure
how to map code section labels to the correct class and specify which
interpreter command to execute to start a session for each class. For example,
to select a Bourne-Again Shell interpreter as \verb|sh|, we add the
\verb|--sh-interpreter=/usr/bin/bash| argument assuming that this binary is
present in the system.

\subsection{Session Management}

\begin{figure*}[hbt!]
  \centering
  \begin{minipage}{\textwidth}
  \begin{verbatim}
  \begin{result}
  ... some output ...
  [LR:/tmp/litrepl/python/partial_c335adc.txt]
  \end{result}
  \end{verbatim}
  \end{minipage}
  \caption{Example partial result section that ends with a continuation tag,
  where \texttt{LR:} is a distinguishable prefix, and the rest is the filename
  storing a response of the interpreter to the section being evaluated.}
  \label{fig:partial-result}
\end{figure*}

Litrepl's ability to maintain interpreter sessions in the background is crucial
for enabling a Read-Eval-Print Loop (REPL) environment. The associated
resources, shown in Figure \ref{fig:resource-allocation}, are stored as files
within an auxiliary directory.

If not specified by command-line arguments or environment variables, the
directory path is automatically derived from the interpreter class name, the
current working directory, and the OS's temporary file location.

The auxiliary directory includes two POSIX pipes for interpreter I/O and a file
recording the running interpreter's process ID, aiding session management.

When a code section is evaluated, Litrepl assigns a response file name derived
from hashing the code. This response file stores the output from the
interpreter.

During evaluation, Litrepl spawns a response reader process with a soft lock,
active until the interpreter completes and responds to an echo probe. The state
machine that operates the probe is the only added hidden state in the entire
system.

If the response exceeds the configured duration, Litrepl outputs a partial
result tag, which is recognized and reevaluated in subsequent runs.  Figure
\ref{fig:partial-result} shows an example partial result section.

Litrepl provides \textbf{start}, \textbf{stop}, \textbf{restart} and
\textbf{status} commands to control background sessions, so, for example

\begin{verbatim}
$ litrepl restart python
\end{verbatim}

stops the Python interpreter if it was running and starts a new instance of it.
The \textbf{interrupt} command sends an interruption signal to the interpreter.
Finally, the \textbf{repl} command establishes direct communication with the
interpreter, allowing manual inspection of its state. The \textbf{help} command
prints the detailed description of each command and the configuration
arguments available.

\subsection{Parsing and Evaluation}

\begin{figure*}[hbt!]
  \centering
  \begin{minipage}{\textwidth}
  \begin{verbatim}
  document       ::= (code | result | ignore | text)*
  code           ::= (code-normal | code-comment)
  result         ::= (result-normal | result-comment)
  code-normal    ::= "\begin{MARKER}" text "\end{MARKER}"
  code-comment   ::= "% MARKER" text "% noMARKER"
  result-normal  ::= "\begin{result}" text "\end{result}"
  result-comment ::= "% result" text "% noresult"
  ignore         ::= "% ignore" text "% noignore"
  text           ::= ...
  \end{verbatim}
  \end{minipage}
  \caption{An illustrative grammar template for \Latex documents where marker
  serves as a parameter configured via command-line arguments for each supported
  interpreter class.}
  \label{fig:document-representation}
\end{figure*}

Litrepl abstracts documents as a straightforward sequence comprising code,
result, and text sections. Additionally, Litrepl identifies ignore blocks, which
act as comments that prevent enclosed sections from being evaluated.

Template grammars similar to the illustrative example in Figure
\ref{fig:document-representation} are encoded for Markdown and \Latex formats.
Before each run, Litrepl calls Lark (\textcite{Lark}) to compile a customized
parser and uses it to access the sections.

Evaluation results are written back into the result sections, and the entire
document is printed. At this stage, certain conditions can be optionally
checked. First, setting \verb|--pending-exitcode| to a non-zero value instructs
Litrepl to report an error if a section takes longer than the timeout to
evaluate. Second, setting \verb|--exception-exitcode| directs Litrepl to detect
Python exceptions. Lastly, \verb|--irreproducible-exitcode| triggers an error if
the evaluation result doesn't match the text initially present in the result
section.

The last option implements the only formal check for aiding reproducibility that
Litrepl provides.

\section{Discussion}

The technical decision to abstract interpreters using text streams comes with
both advantages and disadvantages. A key advantage is simplicity. However, there
are notable negative aspects. First, there is no parallel evaluation at the
communication level, meaning the interpreter is locked until it completes the
evaluation of one snippet before proceeding to the next. Second, the
transferable data type is restricted to text-only streams.

We argue that the lack of parallel execution at the communication level can be
mitigated using interpreter-specific parallelism, where supported. For instance,
Python programs can utilize various subprocess utilities, while shell programs
have full access to shell job control.

The restriction to text-only data types presents a more fundamental limitation.
Litrepl lifts this restriction by supporting text-only document formats. Both
\Latex and Markdown incorporate non-text data without encoding it directly,
instead relying on references and side channels, such as the file system or
network resource identifiers. Consequently, Litrepl shares, for example, the
benefits of human-readable representation in version control systems and the
penalties, such as the need to explicitly organize side-channel data transfer.

Another controversial technical decision is transferring the entire document as
input and output, which can negatively impact performance. Our experience shows
that the system performs adequately for documents of a few thousand lines.
However, larger documents may experience uncomfortable delays, even on modern
computers. Despite this, we choose to maintain this interface because it
simplifies editor integration. A typical plugin can pipe the whole document
through the tool using just a few lines of code.

A more performance-oriented integrations can make pre-parsing and pipe only
relevant document parts. For these approaches, Litrepl offers the
\verb|print-regexp| command, which outputs the anchor regexp in several common
formats.

\section{Conclusion}

The tool is implemented in Python in under 2K lines of code according to the LOC
metric, and has only two Python dependencies so far, at the cost of the
dependency on the POSIX operating system interfaces. Needless to say, we used
Litrepl to evaluate and verify the examples presented in this document.

\printbibliography

\end{document}